\documentclass[a4paper,11pt]{article}
\pdfoutput=1 % if your are submitting a pdflatex (i.e. if you have
\usepackage{jcappub} % for details on the use of the package, please
\usepackage[T1]{fontenc} % if needed

\usepackage{dsfont}
\usepackage{wrapfig}
\usepackage{slashed}
\usepackage{gensymb}

\usepackage{hyperref}
\usepackage[dvipsnames]{xcolor}
\usepackage{amsmath}
\usepackage{amssymb}
\usepackage{mathtools}
\usepackage{siunitx}
\usepackage{feynmp-auto}

\usepackage{cleveref}

\usepackage{float}
\usepackage{subcaption}
\DeclareSIUnit{\parsec}{\text{pc}}

\title{Left-Right Symmetric Neutrino Mass Model without
Scalar Bi-doublet}

\author[a]{Zafri A. Borboruah,}
\author[a]{Lekhika Malhotra,}
\author[b]{Utkarsh Patel,}
\author[b]{Sudhanwa Patra,}
\author[a]{S. Uma Sankar}

\affiliation[a]{Department of Physics, IIT-Bombay, Powai, Mumbai 400076, Maharashtra, India}
\affiliation[b]{Department of Physics, IIT-Bhilai, Durg 491002, Chhattisgarh, India}

\emailAdd{zafri123@iitb.ac.in}
\emailAdd{lekhikamalhotra97@gmail.com}
\emailAdd{utkarshp@iitbhilai.ac.in}
\emailAdd{sudhanwa@iitbhilai.ac.in}
\emailAdd{uma@phy.iitb.ac.in}

\abstract{We consider a left-right symmetric model with an $SU(2)_L$ and an $SU(2)_R$ scalar doublet but without the scalar bidoublet. The charged fermion masses in this model are generated via a universal seesaw mechanism. We add a set of three gauge singlet neutral fermions with Majorana masses of the order of a TeV. The masses for the left-handed neutrinos are naturally small in this model because they occur only at one-loop and are generated through a see-saw mechanism. By an appropriate choice of the Yukawa couplings of the $SU(2)_L$ doublet and the masses of the gauge singlet fermions, it is possible to implement resonant leptogenesis at TeV scale. The right-handed sector of the model, through appropriate values of the Yukawa couplings of the $SU(2)_R$ doublet, leads to a warm dark matter candidate in the lightest right-handed neutrino with a mass of a few keV and an observable effective electron mass for neutrinoless double beta decay $m_{\beta \beta}$. }

\makeatletter
\gdef\@fpheader{}
\makeatother

\begin{document}
\maketitle
\flushbottom

\section{Introduction}

\medskip

The Standard Model (SM) of particle physics has been remarkably successful in describing the fundamental interactions of nature, accurately predicting a wide range of experimental results. However, it remains incomplete, as it fails to explain several crucial observations: the small but nonzero masses of neutrinos, the observed baryon asymmetry of the Universe (BAU), and the nature of dark matter (DM). The discovery of neutrino oscillations in solar, atmospheric, and reactor experiments provides irrefutable evidence that neutrinos have mass, necessitating an extension of the SM framework \cite{Super-Kamiokande:1998kpq,SNO:2001kpb,KamLAND:2002uet,DayaBay:2012fng,T2K:2011ypd}. Moreover, the asymmetry between matter and antimatter in the Universe, quantified by the baryon-to-photon ratio $\eta_B = (6.21 \pm 0.16) \times 10^{-10}$ \cite{Planck:2018vyg}, cannot be generated within the SM alone due to the insufficient CP violation and baryon number-violating interactions at the electroweak scale. Additionally, various astrophysical and cosmological observations indicate that non-luminous, non-baryonic dark matter constitutes a significant fraction of the Universe’s energy density, yet no SM particle possesses the necessary properties to account for it \cite{Rubin:1970zza,Sofue:2000jx}. 

A theoretically well-motivated extension of the SM that addresses these shortcomings is the Left-Right Symmetric Model (LRSM) \cite{Mohapatra:1974gc,Pati:1974yy,Senjanovic:1975rk,Senjanovic:1978ev,Mohapatra:1979ia,Mohapatra:1980yp,Pati:1973uk,Pati:1974vw}. The LRSM restores parity symmetry at high energies, explaining the chiral nature of weak interactions as a consequence of spontaneous symmetry breaking. It naturally accommodates right-handed neutrinos ($\nu_R$), which play a crucial role in generating small neutrino masses via the seesaw mechanism \cite{Minkowski:1977sc,Yanagida:1979as,Yanagida:1980xy,Gell-Mann:1979vob,Mohapatra:1979ia}. Traditional implementations of LRSM contain a scalar bidoublet ($\Phi$) to generate fermion masses via Yukawa interactions. However, the presence of the bidoublet leads to several challenges, such as large tree-level flavor-changing neutral currents (FCNCs) and additional CP-violating phases~\cite{Barenboim:1996rn,Rodriguez:2002ey,Mansour:2014sra,Guadagnoli:2010sd}, which require severe fine-tuning to be phenomenologically viable.

% Several alternative LRSM scenarios have been proposed to address these issues. One approach involves removing the bidoublet while retaining $SU(2)$ triplet scalars to mediate neutrino masses through Type-II seesaw \cite{Patra:2012ur,Borah:2017ldt}. In another variation, the LRSM framework has been explored with inverse and linear seesaw mechanisms, requiring additional singlet fermions to induce small neutrino masses \cite{InverseSeesaw1,LinearSeesaw1,Deppisch:2017vne,Boruah:2022csq}. Some works have considered extended Higgs sectors, including additional triplet or doublet scalars, to modify the neutrino mass structure while preserving the left-right gauge symmetry \cite{Borah:2016hqn,Borah:2017leo}. 

Given the complications associated with the bidoublet scalar, it is natural to ask whether it is truly necessary. In this work, we explore a variant of the LRSM that does not contain the bidoublet or triplet, significantly simplifying the Higgs sector. Instead, we consider a minimal Higgs sector consisting of only an $SU(2)_L$ doublet $H_L$ and an $SU(2)_R$ doublet $H_R$. Without the bidoublet, the usual Dirac mass terms connecting left- and right-handed fermions are absent. However, they can be  generated through a \textit{universal seesaw} mechanism \cite{Davidson:1987mh,Patra:2012ur,Borah:2017ldt,Deppisch:2017vne,Borah:2016hqn,Borah:2017leo, Maharathy:2022gki}, where the charged fermions acquire their masses through their interactions with heavy vector-like fermions. This approach not only avoids tree-level FCNCs but also naturally explains the observed mass hierarchy in the charged fermion sector. 

The absence of the bidoublet and the triplet scalars prevent the standard Type-I seesaw neutrino masses. To generate the neutrino masses, we introduce a set
of three gauge single neutral fermions, with Majorana masses of order ${\cal O} \sim 1$ TeV.
The light neutrino masses arise via a radiative mechanism at the one-loop level, mediated by these singlet fermions~\cite{Cai:2017jrq, Carrasco-Martinez:2023nit,Hall:2023vjb}. This radiative mass generation also incorporates a see-saw mechanism which ensures that the left-handed neutrino masses remain naturally small without requiring extremely high-scale physics. In addition, the coupling of the singlet fermions to right-handed neutrinos generates heavy Majorana neutrino masses through type-I see-saw mechanism. We require the lightest of these heavy neutrinos to have a mass of a few keV, making it a warm dark matter candidate~\cite{Bezrukov:2009th,Nemevsek:2012cd,Kang:2014mea,Kang:2019xuq}. The amplitude for neutrinoless double beta decay, obtained through the mediation of these heavy neutrinos and the right-handed currents, can be large enough to be observable either in the current or in the near future experiments \cite{KamLAND-Zen:2016pfg, Agostini:2022zub}.  It is also possible to obtain sufficient lepton asymmetry in this model via the decay of the singlet fermions through resonant leptogenesis at the scale of $1$ TeV~\cite{Pilaftsis:1998pd,Pilaftsis:2003gt,Dev:2017wwc}. In summary, our model provides a minimalist approach to LRSM without the bidoublet, where the Higgs content is reduced to its simplest viable form while still accommodating neutrino masses, dark matter, neutrinoless double beta decay and leptogenesis within a single framework.

The rest of the paper is structured as follows: In Section~\ref{sec:model}, we explain the details of our model, outlining the fermion content and the scalar sector. This section also discusses generation of the mass for both left and right handed neutrinos. Section~\ref{sec:constraints} discusses the constraints on the model arising from neutrino oscillations, low scale resonant leptogenesis,  keV scale dark matter candidate, observable neutrinoless double beta decay and light-heavy mixing. We also estimate the values of the parameters of the model arising from the above constraints. Section~\ref{sec:numerical} presents our numerical analysis, highlighting the parameter space consistent with experimental data. Finally,
in Section~\ref{sec:discussion}, we summarize our results and discuss possible extensions.

\section{The model}
\label{sec:model}
We consider a left-right symmetric model (LRSM) with a minimal scalar sector, consisting of an $SU(2)_{L}$ doublet $H_{L}$, and an $SU(2)_{R}$ doublet $H_{R}$. Traditional scalar bidoublet and scalar triplets are not present in this scenario. To form $SU(2)_R$ lepton doublets, we introduce three generations of right handed neutrino $\nu_R$. To generate charge fermion masses through universal seesaw, we also introduce vector-like quarks and charged leptons $U_{L,R}, D_{L,R}, E_{L,R}$. In addition, we also introduce a set of three Majorana fermions $S_i$. The fermions of the model have the following gauge quantum numbers~\cite{Dev:2015vjd, Deppisch:2016scs, Patra:2012ur,Deppisch:2017vne},
\begin{gather}
	Q_{L}=\begin{pmatrix}u_{L}\\
	d_{L}\end{pmatrix}\equiv[3,2,1,1/3], \quad Q_{R}=\begin{pmatrix}u_{R}\\
	d_{R}\end{pmatrix}\equiv[3,1,2,1/3]\,,\nonumber \\
	\ell_{L}=\begin{pmatrix}\nu_{L}\\
	e_{L}\end{pmatrix}\equiv[1,2,1,-1], \quad 
	\ell_{R}=\begin{pmatrix}\nu_{R}\\
	e_{R}\end{pmatrix}\equiv[1,1,2,-1] \,, \nonumber \\
    U_{L,R}\equiv[3,1,1,4/3]\,,\quad D_{L,R}\equiv[3,1,1,-2/3]\,, \nonumber \\ E_{L,R}\equiv[1,1,1,-2]\,, \quad S\equiv[1,1,1,0].
\end{gather}
For neutrinos, the conventional type-I seesaw mechanism for mass generation is not possible since there is no Dirac mass term between $\nu_{L,R}$ without the scalar bidoublet and also no Majorana mass for the right-handed neutrinos without the scalar triplet. However, it is possible to couple $\nu_L$ and $\nu_R$ to $S_i$ through $H_L$ and $H_R$ respectively. We will show below that these couplings lead to seesaw masses for right-handed neutrinos and radiatively generated sub-eV masses for left-handed neutrinos.  Additionally, the decay of lightest of $S_i$ can generate adequate lepton asymmetry through resonant leptogenesis. 

%%%%%%%%%%%%%%%%%%%%%%%%%%
\subsection{Scalar masses}
The scalar Lagrangian is given by,
\begin{align}
\mathcal{L}&  = 
   (D_\mu H_L)^\dagger (D^\mu H_L) 
 + (D_\mu H_R)^\dagger (D^\mu H_R) + \mu^2_L |H_L|^2 + \mu^2_R |H_R|^2  \nonumber\\
&- \lambda \left( |H_L|^4 +  |H_R|^4 \right) -\beta |H_L|^2 |H_R|^2 +\mbox{h.c.}\,
\label{eq:slepton_lagrangian}
\end{align}
The left-right symmetry is broken when $H_R\equiv (h_R^+, h^0_R)^T \equiv [1,1,2,1]$ acquires a vacuum expectation value (VEV) while the electroweak symmetry is broken when 
$H_L\equiv (h_L^+, h^0_L)^T \equiv [1,2,1,1]$ acquires VEV.

\begin{align}
    \label{eq:H_vevs}
	\langle H_R \rangle = \begin{pmatrix} 0\\ \frac{v_R}{\sqrt{2}} \end{pmatrix}, \quad 
	\langle H_L \rangle = \begin{pmatrix} 0\\ \frac{v_L}{\sqrt{2}}  \end{pmatrix}. 
\end{align}
By minimizing the scalar potential with respect to the VEVs $v_{L}$ and $v_R$ and expressing $\mu$-parameters in terms of the VEVs, we get the scalar mass matrix,
\begin{equation}\label{eq:MH2}
    M_H^2=\begin{pmatrix} 2\lambda\,v_L^2 & \beta\, v_L\,v_R \\ \beta\, v_L\,v_R & 2\lambda\,v_R^2 \end{pmatrix}.
\end{equation}
The physical scalar masses are obtained by diagonalizing this matrix. In the limit $v_L\ll v_R$ these masses are,
\begin{equation}\label{eq:scalarmasses}
    m_{h_1}^2\simeq 2\lambda\,v_L^2\left(1-\frac{\beta^2}{4\lambda^2}\right),\quad m_{h_2}^2\simeq 2\lambda\,v_R^2.
\end{equation}
% It will be shown in the next section that $\beta =\mathcal{O}(10^{-4})$ to obtain sub-eV light neutrino masses.
Given $m_{h_1} = 126$ GeV and $v_L = 246$ GeV, we find $\lambda= 1/8$ and $m_{h_2} = v_R/2$ assuming $\beta \ll 2 \lambda$.

\subsection{Charged fermion masses via universal seesaw mechanism}
\label{sec3}
In this model, standard Dirac mass terms for Standard Model (SM) fermions are absent due to the lack of a scalar bidoublet. However, by introducing vector-like copies of quark and charged lepton gauge isosinglets, charged fermion mass matrices can be formed with a seesaw structure. The Lagrangian for these vector-like charged fermions is:
\begin{align}
\mathcal{L} = 
	&- (Y_U^L \overline{q}_L \tilde{H}_L  U_R + Y_U^R \overline{q}_R \tilde{H}_R  U_L + Y_L^L \overline{q}_L H_L  D_R \nonumber\\
	&+ Y_L^R \overline{q}_R H_R  D_L + Y_E^L  \overline{\ell}_L H_L E_R + Y_E^R \overline{\ell}_R H_R  E_L) \nonumber\\
	&- M_U \overline{U} U - M_D \overline{D} D - M_E \overline{E} E + \text{h.c.},
\label{2.1}
\end{align}
where, $\tilde{H}_{L,R}$ denotes $i \tau_2 H_{L,R}^\ast$. After spontaneous symmetry breaking, the charged fermion mass matrices become:
\begin{gather}
\label{2.3}
	M_{uU}    = \begin{pmatrix} 0 & Y_U^L v_L \\ Y_U^R v_R & M_U \end{pmatrix}, \,
	M_{dD}    = \begin{pmatrix} 0 & Y_L^L v_L \\ Y_L^R v_R & M_D \end{pmatrix}, \nonumber\\
	M_{eE}    = \begin{pmatrix} 0 & Y_E^L v_L \\ Y_E^R v_R & M_E \end{pmatrix}.
\end{gather}
These matrices generate fermion masses through seesaw mechanism. Assuming real parameters, the SM and heavy vector partner up-quark masses are:
\begin{align}
\label{2.4.0}
	m_u \approx Y^L_U Y^R_U \frac{v_L v_R}{\hat{M}_U}, \quad
	\hat{M}_U \approx \sqrt{M_U^2 + (Y^R_U v_R)^2}.
\end{align}
Similar expressions apply to other fermions. The hierarchy of SM fermion masses can be explained by a hierarchical structure of the Yukawa couplings or the vector-like fermion masses.

\subsection{Gauge boson masses}
The absence of a scalar bidoublet results in zero tree-level mixing between $W_L$ and $W_R$ and hence the charged gauge boson masses are:
\begin{equation}
M_{W_1} = \frac{g}{2} v_L, \quad
M_{W_2} = \frac{g}{2} v_R ,
\end{equation}
where, $g$ is the common gauge coupling for $SU(2)_L$ and $SU(2)_R$. 
% At the one-loop level, a small mixing $\frac{g^2}{16\pi^2} \frac{m_b m_t}{M_{W_R}^2}$ is induced.
Similarly, the neutral gauge boson masses are:
\begin{equation}
M_{Z_1} \approx \frac{g}{2c_W} v_L, \quad
M_{Z_2} \approx \frac{\sqrt{g^2 + g^2_{BL}}}{2} v_R,
\end{equation}
where, $c_W=\cos{\theta_W}$ and $g_{BL}$ is the $U(1)_{B-L}$ gauge coupling. LHC experiments have set the lower limit $M_{W_2} > 6$ TeV \cite{ATLAS:2019lsy} and
$M_{Z_2} > 5.1$ TeV \cite{ATLAS:2019erb, CMS:2021ctt}
on additional charged and neutral gauge bosons. These limits are applicable in our model.  To satisfy the LHC constraints on the masses of $W_R$, the right-handed scalar doublet, $H_R$ should have a VEV $v_R \sim 20$ TeV. Since $M_{Z_2} > M_{W_2}$ in our model, satisfying the experimental constraint on the heavy charged gauge boson mass automatically satisfied that on the heavy neutral gauge boson mass. From Eq.~\eqref{eq:scalarmasses}, we find the mass of the heavy neutral Higgs to be $m_{h_2} = 10$ TeV. In this model, the mixing between the light and heavy gauge bosons is induced only at loop level and is negligibly small.  
%%%%%%%%%%%%%%%%%%%%%%%%%%%%%%%%%%%%%%%%%%%%%%%%%%%%%%

\subsection{Neutrino masses}
\label{sec4}
The Lagrangian for the neutral fermion masses is 
\begin{equation}
    \mathcal{L}=Y_L\overline{\ell_L} \tilde{H}_L S+Y_R\overline{\ell_R^c} H_R S + M_S\overline{S^c}S +h.c.
\end{equation}
where $S$ is assumed to be right-chiral and the generation indices are suppressed for simplicity \cite{Carrasco-Martinez:2023nit, Hall:2023vjb}. After symmetry breaking the doublets acquire VEVs given by Eq.~\eqref{eq:H_vevs}. The tree level mass matrix can be written in the basis $(\nu_L^c,\nu_R,S)$ as,
\begin{align}\label{eq:MnuMatrix}
	M_\nu= 
	\left(\begin{array}{ccc}
		0 & 0 & M_{LS} \\
   	    0 & 0 & M_{RS} \\
        M_{LS}^T & M_{RS}^T & M_S
\end{array} \right) \,.    
\end{align}
where $M_{LS}=Y_L v_L/\sqrt{2}$ and $M_{RS}=Y_R v_R /\sqrt{2}$ are the Dirac mass matrices and $M_S$ is the Majorana mass matrix for singlet fermions $S$. Without loss of generality, we assume $M_S=\text{diag}(M_{S_1},M_{S_2},M_{S_3})$. We also assume the mass hierarchy $M_S>M_{RS} \gg M_{LS}$. To obtain resonant leptogenesis at the scale of $1$ TeV, we take $M_{S_1} = 1$ TeV, $M_{S_2} = 
M_{S_1}\,(1 + \delta)$ with $\delta \ll 1$ and $M_{S_3} = 10$ TeV.

\begin{figure}[ht!]
    \centering
    \includegraphics[width=0.65\linewidth]{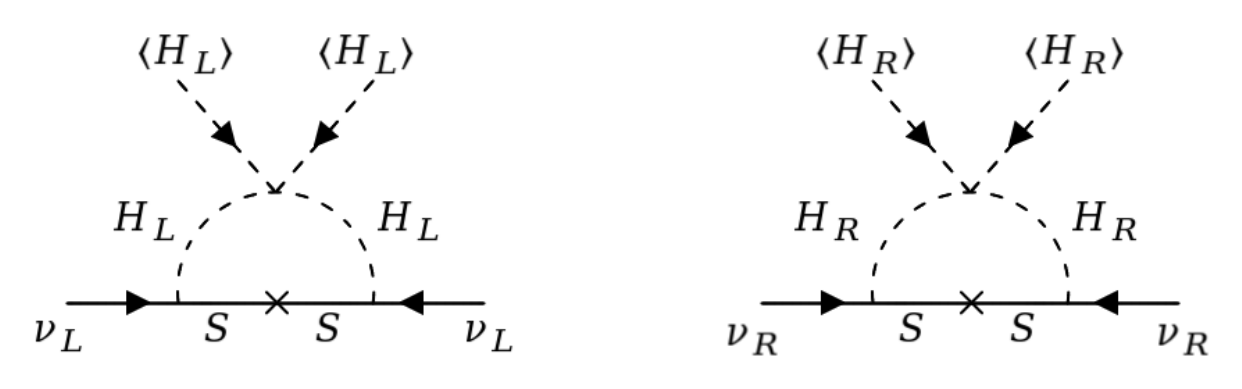}
    \caption{Feynman diagram contributing to neutrino mass at 1-loop.}
    \label{fig:nuMass1loop}
\end{figure}

Diagonalization of the mass matrix in Eq.~\eqref{eq:MnuMatrix} leads to zero masses for left-handed  neutrinos while right-handed neutrinos attain tree level masses via the type I seesaw mechanism. Very light masses for the left-handed neutrinos can be generated at one loop level~\cite{Cai:2017jrq} from the diagram shown on the left in Fig.~\ref{fig:nuMass1loop}. Similar diagram shown on the right in Fig.~\ref{fig:nuMass1loop} contributes a 1-loop level addition to the mass of the right-handed neutrinos. However, it is an order of magnitude smaller than the tree-level mass. 

The leading contributions for the light and heavy neutrino mass matrices are given by
\begin{align}\label{eq:Mnu}
    m_\nu & = \, \frac{v_L^2}{2} Y_{L}\, M_S^{-1} Y_{L}^T I_{\rm loop}\\
    m_N & = -M_{RS} M_S^{-1} M_{RS}^T = - \frac{v_R^2}{2} Y_R M_S^{-1} Y_R^T,
\end{align}
where, $I_{\rm loop}$ is the factor arising from the loop integration and is given by
\begin{equation}
    I_{\rm loop} = \frac{\lambda}{16 \pi^2} \left( \ln\frac{M_{S_i}^2}{m_{h_1}^2} -1  \right).
\end{equation}
%We note that a Dirac mass term between $\nu_L$ and $\nu_R$ is generated at the two-loop level, of the order $m_D \lesssim g_L^4/(16\pi^2)^2 m_\tau m_b m_t / M_{W_2}^2$. This term is negligible for the value of $M_{W_2}$ chosen later.

The complex symmetric light neutrino mass matrix $m_\nu$ is diagonalized by a unitary matrix $U_\nu$ leading to
\begin{align}
    Y_L \, &M_S^{-1} \, \left(\frac{v_L^2}{2} \, I_{\rm loop} \right) \, Y_L^T = m_\nu = U^*_\nu\, \hat{m}_\nu \, U^\dagger_\nu
    \label{eq:casas}
\end{align}
where, $\hat{m}_\nu=\text{diag}(m_1,m_2,m_3)$ is the diagonal light neutrino mass matrix.
The above equation can be rewritten as,
\begin{align}
    \left(v_L \, \sqrt{ I_{\rm loop} / 2}\, \hat{m}_\nu^{-1/2} U_\nu^T \, Y_L \, M_S^{-1/2} \right) \, \left( M_S^{-1/2} Y_L^T U_\nu\, \hat{m}_\nu^{-1/2}\, v_L \, \sqrt{ I_{ \rm loop} / 2}\right) = \mathbb{I}.
\end{align}
From this Casas-Ibarra form~\cite{Ibarra:2011xn,Lopez-Pavon:2015cga,Li:2024uat} we define the complex orthogonal matrix 
\begin{equation}
    \mathcal{O}_L = v_L \, \sqrt{\, I_{ \rm loop} / 2}\, \hat{m}_\nu^{-1/2} U_\nu^T \, Y_L \,M_S^{-1/2}.
    \label{eq:CasasIbarra}
\end{equation}
The Yukawa matrix $Y_L$ can be written in terms of the known light neutrino masses and mixings and the unknown heavy fermion masses and $\mathcal{O}_L$ as,
\begin{align}\label{eq:casasYL}
  Y_L =  \frac{1}{v_L}\,\sqrt{\frac{2}{I_{\rm loop}}} U_\nu^* \sqrt{\hat{m}_\nu} \, \mathcal{O}_L \, \sqrt{M_S}
\end{align}
%Leptogenesis imposes some constraints on the structure of the Yukawa matrix $Y_L$ which we discuss in the next sub-section.

\section{Constraints of the parameters of the model} \label{sec:constraints}
The model has three mass scales: $v_L,v_R$ and $M_S$. The neutrino masses and mixings of the model also depend on the two Yukawa matrices $Y_L$ and $Y_R$.  In this section, we systematically discuss the various measurable physical quantities that can be obtained from this model and the constraints these quantities impose on the parameters of the model. 

\subsection{Resonant low-scale leptogenesis}\label{sec:lowscale}
\begin{figure}[!ht]
    \centering
    \includegraphics[width=0.8\linewidth]{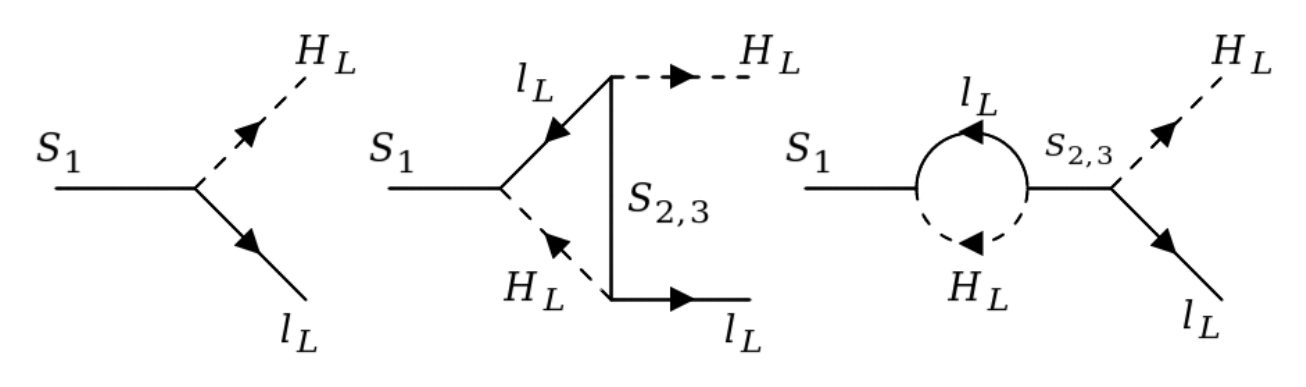}
    \caption{Feynman diagrams contributing to the CP asymmetry parameter $\varepsilon_1$.}
    \label{fig:feynman1}
\end{figure}

The leptogenesis in our model arises through the decay of the lightest singlet fermion $ S_1 $. The CP asymmetry 
\begin{equation}
    \varepsilon_1 = \frac{\Gamma(S_1 \to \ell_L H_L) - \Gamma(S_1 \to \ell_L^\dag H_L^\dag)}{\Gamma_{\text{tot}}(S_1)},
\end{equation}
is calculated from the interference of tree-level and loop-level diagrams shown in Fig.~\ref{fig:feynman1}. This asymmetry arises only through the decay of $S_1$ into particles with $SU(2)_L$ interactions. The neutral fermion $S_1$ does couple to particles with $SU(2)_R$ interactions (such as $\ell_R$ and $H_R$) but the mass of the heavy Higgs scalar $h_2 \approx H_R$ ($m_{h_2} = 10$ TeV) is larger than mass of $S_1$ ($M_{S_1} = 1$ TeV) in our model. Therefore, the  decay $S_1 \to \ell_R H_R$ is forbidden and the coupling of $S_1$ to right-handed sector does not lead to any lepton asymmetry.

The CP asymmetry can be expressed as
\begin{equation}\label{eq:epsilon_1}
    \varepsilon_1 = \frac{1}{8\pi} \sum_{k = 2,3} \left( g_v(x_k) + g_s(x_k) \right) \mathcal{T}_{k1},
\end{equation}
where $ x_k = M_{S_k}^2 / M_{S_1}^2 $. The loop factors are $ g_v(x_k) = \sqrt{x_k}\{1 - (1+x_k) \ln[(1+x_k)/x_k]\} $ and $ g_s(x_k) = \sqrt{x_k}/(1-x_k) $ and 
\begin{equation}
    \mathcal{T}_{k1} = \frac{\text{Im}[(Y_L^\dag Y_L)_{k1}^2]}{(Y_L^\dag Y_L)_{11}}.
\end{equation}

For $ x_k = \mathcal{O}(1) $, the self-energy contribution dominates. For the values of $M_{S_i}$ we have chosen, the CP asymmetry simplifies to
\begin{equation}\label{eq:ep_res}
    \varepsilon_1 \simeq -\frac{1}{16\pi} \left[ \frac{M_{S_2}}{v_L^2} \frac{\text{Im}[(Y_L^\dag m_\nu Y_L^*)_{11}]}{(Y_L^\dag Y_L)_{11}}   \right] R,
\end{equation}
where $R$ is the resonant factor given by $M_{S_1}/(M_{S_2} - M_{S_1}) = (1/\delta)$. To obtain a large enough CP asymmetry, a resonance factor of aboout
$R = 10^6$ is needed. Hence, we take $\delta = 10^{-6}$ in our numerical analysis.

The decay width of $S_1$ is given by,
\begin{equation}
    \Gamma_{\text{tot}}(S_1) = \frac{(Y_L^\dag Y_L)_{11}}{4\pi} M_{S_1}.
    %\approx \frac{(Y_R Y_R^\dag)_{11}}{4\pi} M_{S_1},
\end{equation}
% where, we assumed $(Y_{L} Y_L^\dag)_{11} \ll (Y_{R} Y_R^\dag)_{11}$. 
The out-of-equilibrium condition, $ \Gamma_{\text{tot}}(S_1) < \mathcal{H}(T \sim M_{S_1}) $, imposes an upper limit on $ (Y_L)_{i1} $:
\begin{equation}\label{eq:out-of-eq-cond}
    \sqrt{\sum_i |(Y_L)_{i1}|^2} < 3 \times 10^{-7} \left( \frac{M_{S_1}}{\text{ TeV}} \right)^{1/2}.
\end{equation}
As stated earlier, we fix $M_{S_1}=1$ TeV and $M_{S_2}$ to be almost degenerate so that adequate lepton asymmetry can be generated through resonant leptogenesis. The value of $M_{S_3}$ should be somewhat larger, which we take to be $10$ TeV. 
For $M_{S_1}=1$ TeV, the constraint in Eq.~\eqref{eq:out-of-eq-cond} implies that 
\begin{equation}
|(Y_L)_{i1}| \sim 10^{-7}.
\label{eq:YL1iconst}
\end{equation}
This constraint on the first column of $Y_L$, together with the tiny masses for the left-handed neutrinos, sets very strong constraints on the complex orthogonal matrix ${\cal O}_L$, defined in
Eq.~\eqref{eq:CasasIbarra}. 
The matrix ${\cal O}_L$ can be paramterized as following~\cite{Casas:2001sr},
\begin{equation}\label{eq:OLparameterization}
    \mathcal{O}_L=\begin{pmatrix}
        \hat{c}_2\hat{c}_3 & -\hat{c}_1\hat{s}_3 - \hat{s}_1\hat{s}_2\hat{c}_3 & \hat{s}_1\hat{s}_3 - \hat{c}_1\hat{s}_2\hat{c}_3 \\
\hat{c}_2\hat{s}_3 & \hat{c}_1\hat{c}_3 - \hat{s}_1\hat{s}_2\hat{s}_3 & -\hat{s}_1\hat{c}_3 - \hat{c}_1\hat{s}_2\hat{s}_3 \\
\hat{s}_2 & \hat{s}_1\hat{c}_2 & \hat{c}_1\hat{c}_2
    \end{pmatrix},
\end{equation}
where $\hat{c}_i=\cos\theta_i$ and $\hat{s}_i=\sin\theta_i$, with $\theta_i$ ($i=1,2,3$) are complex angles. The strong constraint in Eq.~\eqref{eq:YL1iconst} on the magnitudes of $(Y_L)_{i1}$, in turn, severely limits the magnitudes of the complex angles $\theta_i$. 
%From our numerical analysis, we find that imposing the constraint $|\theta_i| \leq 1.4^\circ$ satisfies the upper bound on $(Y_L)_{1i}$.

\subsection{Light-heavy neutrino mixing}

The diagonalization of $M_\nu$, given in Eq.~\eqref{eq:MnuMatrix}, leads to light-heavy mixing of order,
\begin{equation}\label{eq:Umixing}
   \left( M_{LS}M_{RS}^{-1}\right) \simeq (Y_L v_L) \, (Y_R \, v_R)^{-1}\equiv U_{LH}.
    \end{equation}
This mixing makes the PMNS matrix deviate from unitarity. This deviation is parameterized as $U_{PMNS}=(1-\eta)U_\nu$, where $U_\nu$ is the diagonalizing unitary matrix of the light neutrino mass matrix $m_\nu$ and $\eta\sim U_{LH}U^\dagger_{LH}$~\cite{Korner:1992zk,Grimus:2000vj}.
The latest flavor and electroweak precision data lead to the following bounds on the elements of matrix $\eta$~\cite{Fernandez-Martinez:2016lgt,Blennow:2023mqx,Fernandez-Martinez:2024bxg},
% $|\eta_{\alpha\beta}|\lesssim 10^{-5}-10^{-3}$~\cite{refs}.
\begin{equation}\label{eq:non unitarity bound}
    |\eta|\leq\begin{pmatrix}
        1.3\times10^{-3} & 1.2\times10^{-5} & 1.4\times10^{-3}\\
        1.2\times10^{-5} & 2.2\times10^{-4} & 6\times10^{-4}\\
        1.4\times10^{-3} & 6\times10^{-4} & 2.8\times10^{-3}
    \end{pmatrix}.
\end{equation}
The elements of $Y_L$ have magnitudes $\sim 10^{-7}$ and we will argue in the next sub-section that the elements of $Y_R$ have magnitudes greater than $10^{-5}$. Hence, the elements of $U_{LH}$ have magnitudes less than $10^{-4}$ and elements of $\eta$ have magnitudes less than $10^{-8}$. Therefore, the constraints on the deviation of PMNS matrix from unitarity are trivially satisfied.

\subsection{Warm dark matter candidate}

The Majorana mass matrix of the right-handed neutrinos is given by 
\begin{equation}
    m_N = -\frac{v_R^2}{2} Y_R (M_S)^{-1} Y_R^T.
    \label{rhnumass}
\end{equation}
This complex symmetric matrix is diagonlised by the unitary matrix $U_N$ to give
\begin{equation}
    U_N^T \, m_N \, U_N = \hat{m}_N = \text{diag}\,(m_{N_1},m_{N_2},m_{N_3})
    \end{equation}
We take $m_{N_1}$ to be the smallest eigenvalue and require it to be of the order of keV and the corresponding eigenstate $N_1$ is a warm dark matter candidate. This state $N_1$ can decay into three light neutrinos through
light-heavy mixing via off-shell $Z_1$ boson exchange. The decay rate can be estimated as,
\begin{equation}\label{eq:gamma3nu}
     \Gamma(N_1\to 3\nu)\sim \frac{G_F^2 m_{N_1}^5}{192\pi^3}\eta_{ee}\simeq \Gamma_\mu \,\left(\frac{m_{N_1}}{m_\mu }\right)^5\,\eta_{ee},
\end{equation}
where $\Gamma_\mu$ is the decay rate for the muon. For $\eta_{ee} < 10^{-8}$ and $m_{N_1}$ in the range $(0.5,10)$ keV, the lifetime of $N_1$ is in the range $(10^{17},10^{23})$ years, much larger than the age of the Universe. 

We also require the other two eigenvalues $(m_{N_2} \, {\rm and} \, m_{N_3})$  to be greater than $200$ MeV, so that the states $N_2$ and $N_3$ can decay into $N_1$ and a pair of oppositely charged leptons. To obtain such eigenvalues with a wide splitting between the lowest value and the other two, we need to take the matrix $Y_R$ to be of the form 
\begin{equation}
    Y_R=\begin{pmatrix}
        \kappa\, a_1 & b_1 & c_1\\
        \kappa\, a_2 & b_2 & c_2\\
        \kappa\, a_3 & b_3 & c_3
    \end{pmatrix}, \label{eq:YRform}
\end{equation}
where $\kappa$ is small number which we choose to be $10^{-2}$ and $a_i,b_i,c_i$ are uniform random complex numbers with magnitudes of the order of $0.01$.

In the mass basis of the light and heavy neutrinos, their charged current interactions can be written as 
\begin{equation}
    {\cal L}_{\rm c.c.} = \frac{g}{\sqrt{2}} \left[ \sum_i (U_\nu)_{\alpha i} \, \bar{\ell}_{\alpha L}
    \gamma^\mu \nu_{i L} \, W^-_{L\mu} + 
    \sum_i (U_N)_{\alpha i} \, \bar{\ell}_{\alpha R}
    \gamma^\mu \nu_{i R} \,W^-_{R\mu} \right] + h.c.
    \label{eq:Lcc}
\end{equation}
The neutrinos $N_{2,3}$ are massive enough to decay into a charged lepton $\ell_\alpha^+\, (\alpha = e,\,\mu)$ and an off-shell $W_R^-$, which, in turn, decays into another charged lepton $\ell_\beta^-\,(\beta=e,\,\mu)$ and $N_1$. The rate for this decay can be estimated as
\begin{eqnarray} 
    \Gamma (N_{2,3} \to \ell^+_\alpha\,\ell^-_\beta\,N_1) & = & \frac{G_F^2 m_{N_{2,3}}^5}{192 \pi^3} \left(
    \frac{v_L}{v_R} \right)^4 |(U_N)_{\alpha 2, \alpha 3}\, (U_N)_{\beta 1}|^2 \, F_{\rm p.s.} \nonumber \\
    & = & 
    \Gamma_\mu \left( \frac{m_{N_{2,3}}}{m_\mu} \right)^5
    \left( \frac{v_L}{v_R} \right)^4
|(U_N)_{\alpha 2, \alpha 3}\, (U_N)_{\beta 1}|^2 \, F_{\rm p.s.} 
\end{eqnarray}
where $F_{\rm p.s.}$ is the phase space factor for the decay. Given $(v_L/v_R) \simeq 10^{-2}$ and $(m_{N_{2/3}}/m_\mu) \gtrsim 2$, we find that the
the lifetime of $N_{2,3}$ will be a few minutes, assuming that neither the phase factor $F_{\rm p.s.}$ nor the mixing matrix elements $|(U_N)_{\alpha 2,\alpha 3}\, (U_N)_{\beta 1}|^2$ are too small.

\subsection{Neutrinoless double beta decay}

\begin{figure}[!ht]
    \centering
    \includegraphics[width=0.6\linewidth]{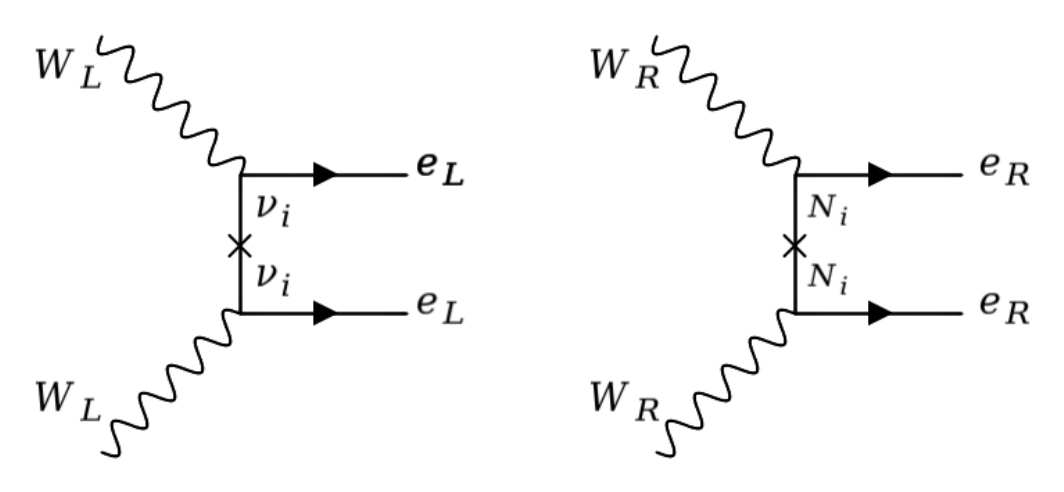}  
    \caption{Feynman diagrams contributing to neutrinoless double beta decay.}
    \label{fig:0nubb}
\end{figure}

The model contains Majorana masses for both the light (left-handed) neutrinos $\nu_i$ and the heavy (right-handed) neutrinos $N_i$. Hence, neutrinoless double beta decay can take place through the exchange of $\nu_i$ as well as that of $N_i$. In Fig.~\ref{fig:0nubb}, the $\nu_i$ exchange diagram is shown on the left and the $N_i$ exchange diagram is shown on the right. 
The amplitude for the left diagram ${\cal A}_L$ and that for the right diagram ${\cal A}_R$ are given by 
\begin{eqnarray}
    {\cal A}_L & \propto &  G_F^2 \frac{\sum_i (U_\nu)_{ei}^2 m_i}{p^2} \equiv G_F^2 \frac{m_{LL}}{p^2} \nonumber \\
    {\cal A}_R & \propto & G_F^2 \left( \frac{v_L}{v_R} \right)^4 \sum_i \frac{(U_N)_{ei}^2 m_{N_i}}{p^2 + m_{N_i}^2} \equiv G_F^2 \frac{m_{RR}}{p^2}
    \label{eq:0nubbamp}
\end{eqnarray}
where $p^2$ is the momentum exchange in the neutrinoless double beta decay process. In our calculations, we take $p = 100$ MeV. The effective electron mass for neutrinoless double beta decay, $m_{\beta \beta}$, is the sum of the effective masses $m_{LL}$ and $m_{RR}$ defined in Eq.~\eqref{eq:0nubbamp}. KamLAND-Zen experiment has set an upper bound of $0.2$ eV on this mass~\cite{KamLAND-Zen:2016pfg}. We find that our choice of parameters leads to $m_{LL}$ being negligibly small and $m_{RR}$ dominating $m_{\beta \beta}$.

%%%%%%%%%%%%%%%%%%%%%%%%%%%%%%%%%%%%%%%%%%%%%%%%%%%%%%%%%%%%%%%%

\section{Numerical analysis}
\label{sec:numerical}
In this section, we describe the numerical procedure we used to search for the allowed values of the parameter space, which satisfy all the constraints from 
\begin{enumerate}
    \item neutrino oscillations
    \item low-scale resonant leptogenesis
    \item mass of warm dark matter candidate
    and the masses of the heavier right-handed neutrinos
    \item neutrinoless double beta decay.
\end{enumerate}

The value of $v_L$ is $246$ GeV and, as mentioned earlier, we fix $v_R = 20$ TeV, $M_{S_1} = 1$ TeV,
$M_{S_2} = M_{S_1} (1+\delta)$ and $M_{S_3} = 10$ TeV.
We choose appropriate random values for the elements of the Yukawa matrices $Y_L$ and $Y_R$. For each choice, we verify that all the constraints listed above are satisfied. 

The constraints from neutrino oscillations are trivially satisfied by choosing $Y_L$ to be of the form given in Eq.~\eqref{eq:casasYL},
\begin{align}\label{eq:casasYL2}
  Y_L =  \frac{1}{v_L}\,\sqrt{\frac{2}{I_{\rm loop}}} U_\nu^* \sqrt{\hat{m}_\nu} \, \mathcal{O}_L \, \sqrt{M_S}.
\end{align}
The matrix $Y_L$ is fully determined in terms of $m_i$, $(U_\nu)_{\alpha i}$, $M_{S_i}$, $v_L$, $I_{\rm loop}$ and the unknown complex orthogonal matrix $\mathcal{O}_L$. We limit our discussion to normal hierarchy (NH) and fix the value of the lightest neutrino mass $m_1$ to be $10^{-6}$ eV. With such light mass, the cosmological constraint on the sum of light neutrino masses, $\sum_i m_i<0.12$ eV~\cite{Planck:2018vyg,eBOSS:2020yzd} is trivially satisfied. To construct the matrices $\hat{m_\nu}$ and $U_\nu$, we use the central values of the neutrino oscillation parameters from the global fit, given in ref.~\cite{Esteban:2024eli}.
These parameters are listed in Table~\ref{tab:nuOsc}. 

\begin{table}[ht!]
\vspace{1mm}
\centering
\begin{tabular}{|c|c|}
\hline \hline
 {\textbf{Oscillation}}   &  {\textbf{Numerical Input}}\\
 %\cline{2-3}
 {{\bf parameters}} & {\tt (NH assumed)}   \\ \hline \hline
 $\Delta m_{21}^2 (10^{-5}~{\rm eV}^2$)  &    7.49    \\ 
 % \hline
  %$\Delta m_{23}^2 (10^{-3}~{\rm eV}^2) $(IH) &   2.414 - 2.581  &     \\ 
 $\Delta m_{31}^2 (10^{-3}~{\rm eV}^2) $ &     2.534  \\ 
  $\sin^2{\theta_{12}}$   &      0.307   \\ 
  %$\sin^2{\theta_{23}}$  (IH) &   0.419 - 0.617  &         \\
 $\sin^2{\theta_{23}}$   &     0.561   \\ 
   %$\sin^2{\theta_{13}} $  (IH) &   0.02052 - 0.02428  &     \\
  $\sin^2{\theta_{13}}  $  &      0.02195    \\ 
   %$\delta_{CP}/^\circ$ (IH) & 193 - 352 &     \\ 
  $\delta_{CP}/^\circ$  &  177  
  %%%%%%%
  \\\hline \hline
\end{tabular}
\caption{Values of neutrino oscillation parameters used as inputs in our numerical analysis, which are the central values from a recent global fit~\cite{Esteban:2024eli}. We have considered only normal hierarchy (NH).}
\label{tab:nuOsc}
\end{table}

 We note that, for the value of $m_1$ considered here, the value of $m_{LL}$ in neutrinoless double decay is negligible. Hence, in our model, the effective electron mass $m_{\beta \beta}$ is dominated by $m_{RR}$ which arises from the right-handed sector.  
 
\begin{figure}[!ht]
    \centering
    \includegraphics[width=0.6\linewidth]{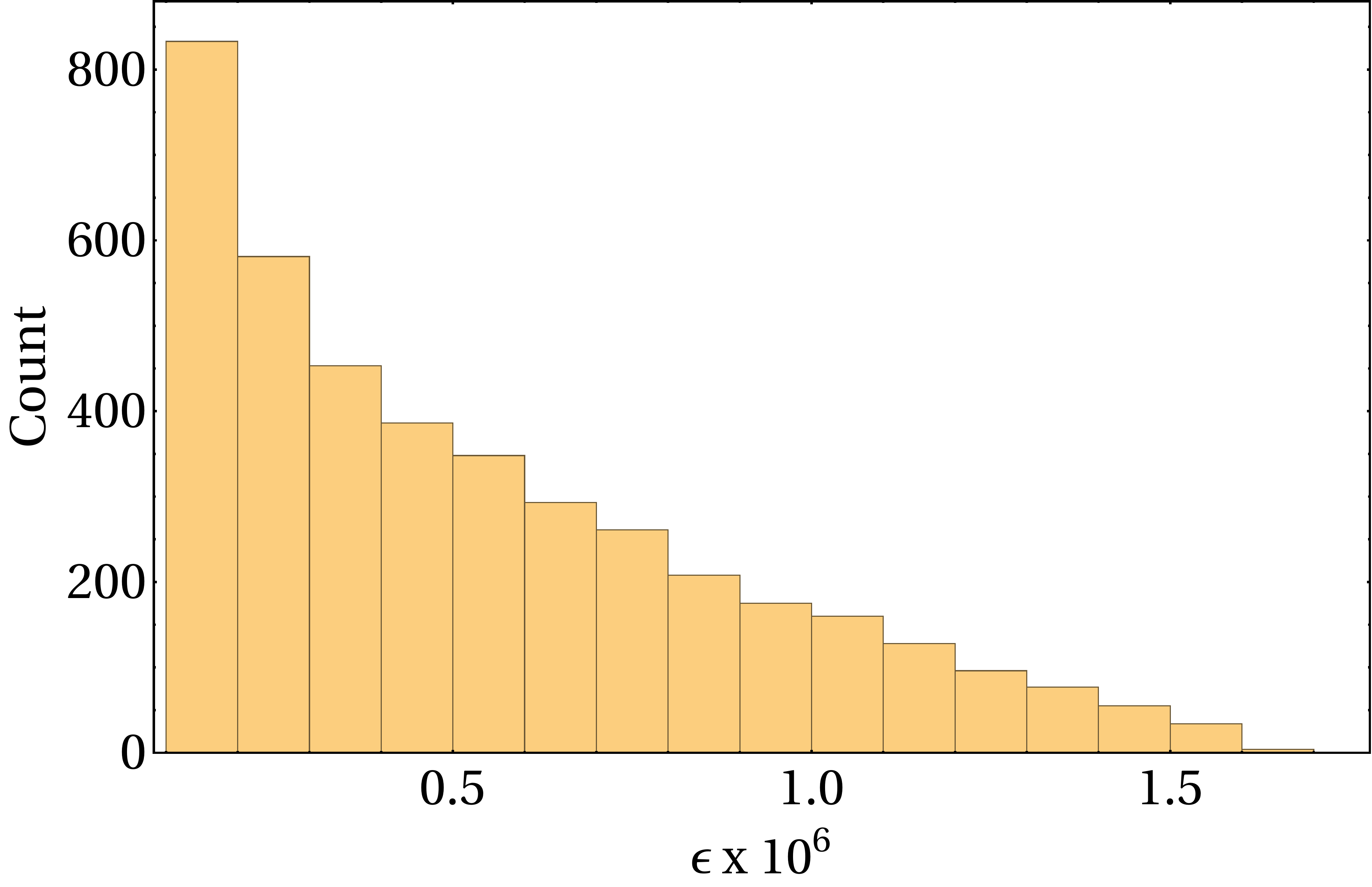} 
    \caption{Histogram of the distribution of the values of $\varepsilon_1$. Total number of values is about $4,000$.}
    \label{fig:epsilon1}
\end{figure}

 To obtain the matrix ${\cal O}_L$, we choose the mixing angles $\theta_i$ to be random complex variables with magnitudes in the range $(0,1^\circ)$. We also require the resonance factor $R$ to be $10^6$ which implies that 
$\delta = 10^{-6}$. We generate $10,000$ sets of random values for $\theta_i$ and construct the corresponding matrices ${\cal O}_L$ and $Y_L$. Of these $10,000$ matrices, we select those which satisfy the following conditions
\begin{equation}
    |(Y_L)_{i1}| \leq 10^{-7}~~~~{\rm and}~~~~
    \varepsilon_1 > 10^{-7}.
    \end{equation}
Of the $10,000$ trials, we find that these conditions are satisfied in about $4,000$ ($40\%$) cases. The values of $\varepsilon_1$ are in the range $(10^{-7}, 2\times 10^{-6})$ in the selected cases. The histogram of the distribution of values of $\varepsilon_1$ is shown in Fig.~\ref{fig:epsilon1}.

\begin{figure}[!ht]
    \centering
    \includegraphics[width=0.6\linewidth]{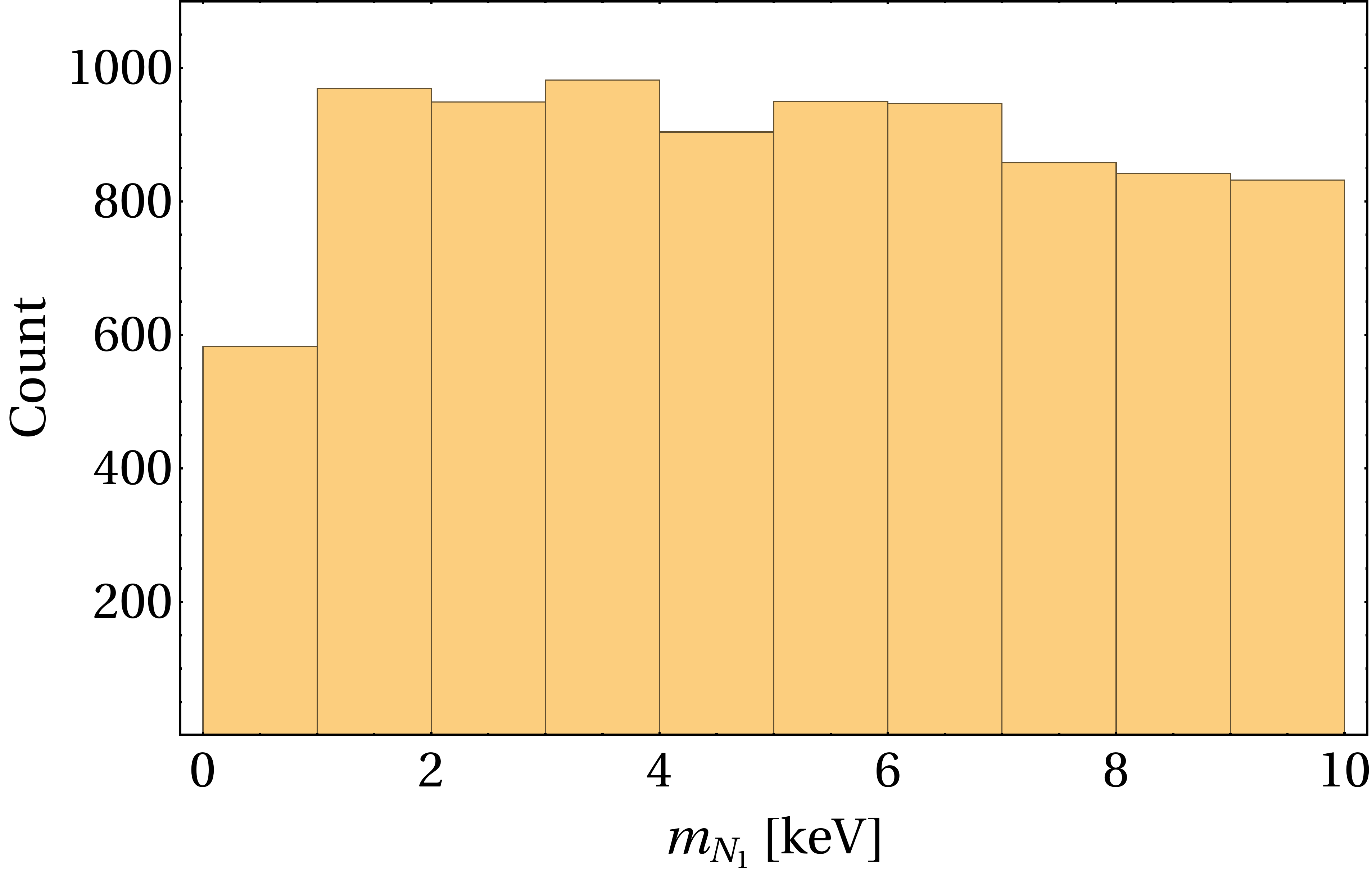} 
    \caption{Histogram of the distribution of the values of $m_{N_1}$. Total number of values is about $9,000$.}
    \label{fig:mN1}
\end{figure}

As discussed in the previous section, we choose the matrix $Y_R$ to be of form given in Eq.~\eqref{eq:YRform}. To obtain $m_{N_1}$ in keV range and $m_{N_{2,3}}$ above $200$ MeV, we choose $\kappa = 10^{-2}$ and $a_i$, $b_i$ and $c_i$ to complex uniform random numbers with magnitudes in the range $(0.005,0.05)$. We choose $5\times10^6$ sets of uniform complex random numbers as the elements of $Y_R$, construct the corresponding $m_N$ matrix and evaluate its eigenvalues and its diagonalising  matrix $U_N$. We impose the following constraints on the eigenvalues and the elements of $U_N$
\begin{equation}
    m_{N_1} \in (0.5,10)\,{\rm keV}; \, m_{N_{2,3}} > 200\,{\rm MeV}\,\,\,{\rm and}\,\,\, 0.01 \leq \, m_{RR} \leq 0.2\,{\rm eV}.
\end{equation}
We find that, of the $5\times10^6$ trials, only about $9,000$ possibilities (about $0.2\%$) satisfy all the constraints above. The histogram of the values of the warm dark matter candidate mass is shown in Fig.~\ref{fig:mN1}. 

\begin{figure}[!ht]
    \centering
    \includegraphics[width=\linewidth]{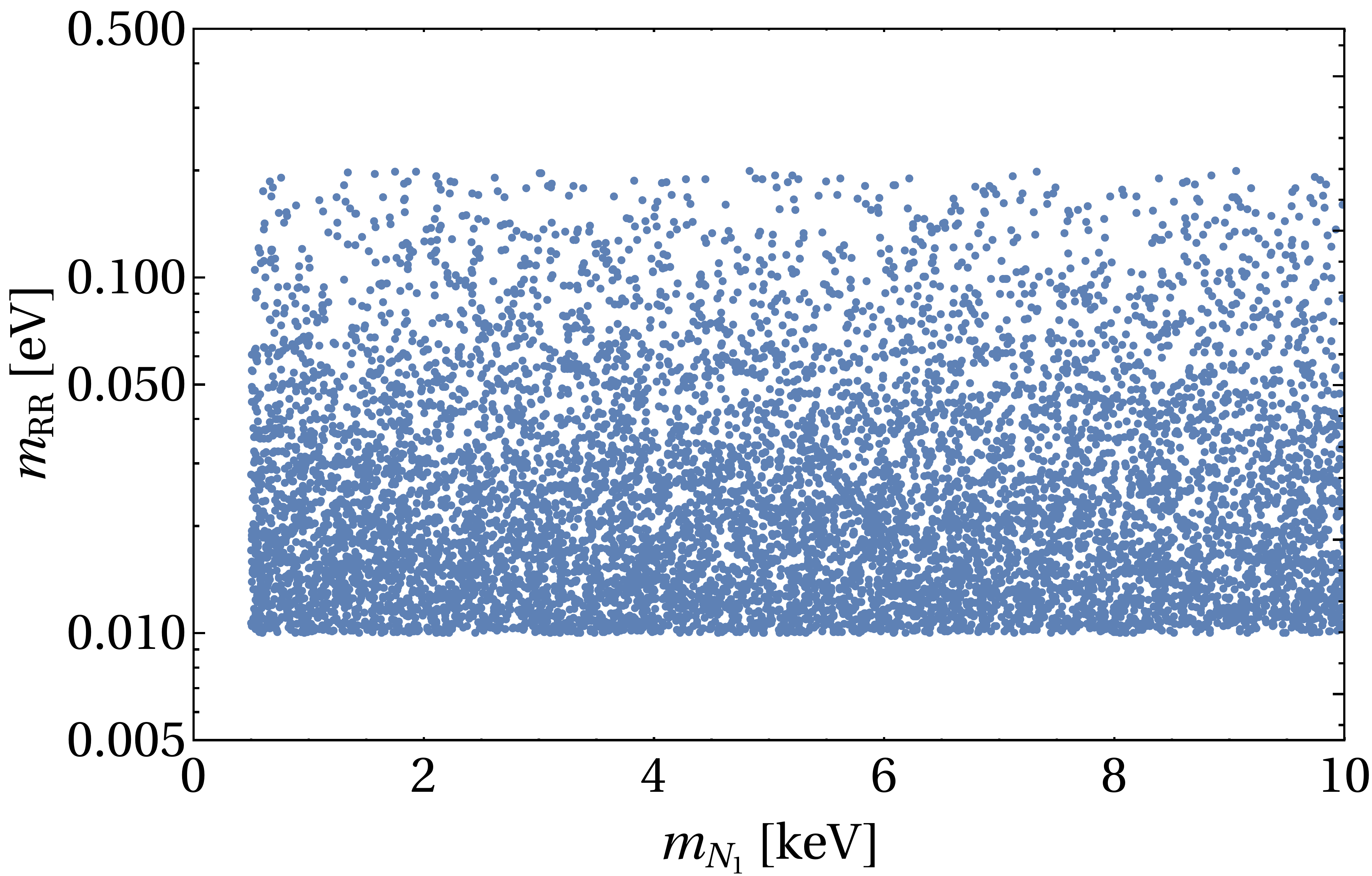} 
    \caption{Scatter plot of allowed points in the 
    $m_{N_1}-m_{RR}$ plane. Total number of points is about $9,000$.}
    \label{fig:mN1vsmee}
\end{figure}

The scatter plot of the allowed values of $m_{N_1}$ vs 
$m_{RR} = m_{\beta \beta}$ is shown in Fig.~\ref{fig:mN1vsmee}. From this plot, we see that for a small set of allowed values of $Y_R$, the value of $m_{\beta \beta}$ is in the range $(0.01,0.2)$, large enough to be observable in the current or planned experiments.

%We define a test point to be a given set of randomly chosen values of $a_i,b_i$, $c_i$ and the angles $\theta_1,\theta_2$ and $\theta_3$. For each test point, we calculate the masses of the three sterile neutrinos, $m_{N_1},m_{N_2}$ and $m_{N_3}$, and the CP asymmetry parameter $\varepsilon_1$. We retain the test point if the following constraints are satisfied:

\section{Discussion}\label{sec:discussion}
We have constructed a neutrino mass model based on left-right symmetry which does not contain a scalar bi-doublet or scalar triplets. Very small light neutrino masses arise in this model naturally through a combination of loop effects and see-saw mechanism. This model satisfies all the constraints related to light neutrino masses, neutrino oscillations, neutrinoless double beta decay and active-sterile mixing. Through numerical analysis, we demonstrated that the model (a)
contains a keV-scale right-handed neutrino which is a dark matter candidate and
(b) can generate necessary CP asymmetry for TeV-scale resonant leptogenesis. 
Thus, our model accounts for all the desired properties related to neutrinos. 
%It is desirable to calculate the dark matter number density in this model and compare it to the observations. Another interesting question is: how leptogenesis is modified if the mass scale of $M_{S_1}$ is increased by two orders of magnitude so that $S_1$ can decay into particles with $SU(2)_R$ interactions also. These questions will be addressed in a future work.

%A long-lived keV-scale sterile neutrino is typically overproduced in the early Universe~\cite{Drewes:2016upu}. However, their abundance can be diluted by injecting entropy into the Universe from late time decays of heavier particles~\cite{Scherrer:1984fd}. Thus, all the required ingredients to satisfy the relic density bounds of dark matter are present within our model. Successful dilution imposes stringent constraints on the diluter's mass, couplings, and lifetime, ensuring decay occurs after neutrino freeze-out but before Big Bang Nucleosynthesis (BBN). The resulting parameter space is highly restrictive, and a detailed analysis of these constraints is left for future work.

\section*{Acknowledgements}
ZAB is supported by DST-INSPIRE fellowship and LM is supported by a UGC fellowship. ZAB and SUS thank the Institute of Eminence (IoE) funding from the Government of India. UP acknowledges the financial support from the Indian Institute of Technology, Bhilai, and the Ministry of Education, Government of India, for conducting the research work. SP would like to acknowledge the funding support from SERB, Government of India, under MATRICS project with grant no. MTR/2023/000687.
%%%%%%%%%%%%%%%%%%%%%%%%%%%%%%%%%%%%%%%%%%%%%%%%%%%%%%%%%%%%%%%%%%%
 
\bibliographystyle{JHEP}
\bibliography{ref}
\end{document}